\documentstyle[prl,aps,epsf,multicol]{revtex}
\begin{document}

\draft

\title{Centrifugal terms in the WKB approximation and semiclassical
quantization of hydrogen}
\author{Joachim Hainz and Hermann Grabert}
\address{Fakult\"at f\"ur Physik, Albert--Ludwigs--Universit{\"a}t, \\
Hermann--Herder--Stra{\ss}e~3, D--79104 Freiburg, Germany}

\date{\today}
\maketitle
\widetext

\begin{abstract}
A systematic semiclassical expansion of the hydrogen problem about the
classical Kepler problem is shown to yield remarkably accurate
results. {\it Ad hoc} changes of the centrifugal term, such as the standard 
Langer modification where the factor $l(l+1)$ is replaced by
$(l+1/2)^2$, 
are avoided. The semiclassical energy levels are shown 
to be exact to first order in $\hbar$ with all higher order contributions 
vanishing. The wave functions and dipole matrix elements are also discussed.
\end{abstract}


\raggedcolumns
\begin{multicols}{2}
\narrowtext 
While the solution of the hydrogen problem was one of the early
successes of quantum mechanics, it failed to be a showpiece for the WKB
approximation which proved to do rather poorly. Usually, this is
attributed to the singularity of the Coulomb potential at $r=0$ where
$r$ is the distance between proton and electron. Clearly, near the
origin the WKB expansion cannot be justified even in the semiclassical limit. 
Langer \cite{Langer} has shown that the correct behavior for 
$r\rightarrow 0$ can be enforced to the WKB wave function if the centrifugal 
term
\begin{equation}
V_{C}(r)=\frac{\hbar^2 l \left(l+1\right)}{2 m r^2}
\label{eq:Langerterm2}
\end{equation}
in the radial Schr\"odinger equation is replaced by
\begin{equation}
V_{L}(r)=\frac{\hbar^2 \left(l+\frac{1}{2}\right)^2}{2 m r^2}.
\label{eq:Langerterm3}
\end{equation}
Quite remarkably, with the Langer modification (LM) (\ref{eq:Langerterm3})
of the interaction potential, the WKB approximation gives exact energy
eigenvalues for the hydrogen problem already to lowest order. As a
consequence, the LM is now seen as a standard
ingredient of WKB theory for the hydrogen problem and related systems
with radial symmetry, such as the radial harmonic oscillator or the
Morse potential in three dimensions. For recent applications and
extensions we refer to the work by Yi {\it et al.} \cite{Yi} and by
Morehead \cite{Morehead}.

 In the last years some attempts have been made to avoid the Langer
 modification. For the exactly solvable hydrogen problem semiclassical
 theories based on nonlinear transformations \cite{Duru} or
 supersymmetry \cite{super} are powerful alternatives to conventional WKB
 methods. However, these approaches lead to exact results only for the
 strict $1/r$ potential and do not constitute a general replacement
 for the standard semiclassical expansion. Within the conventional
 approach, Friedrich and Trost \cite{Friedrich} have avoided the
 LM, introducing instead an additional phase of the WKB wave function which is
 then optimized. For the repulsive $1/r^2$ potential their method give
 results that are superior to those derived from conventional WKB with
 LM. However, also their approach maintains that for
 Coulomb type problems the textbook WKB expansion needs to be modified. 
In this Letter we challenge this common believe.

We start from the obvious observation that in the classical limit the
hydrogen problem should reduce to the Kepler problem. The form of the
classical orbits depends on the energy $E$ and the angular momentum
$L$. Hence, the leading order WKB radial wave function should also be
calculated for given $E$ and $L=\hbar l$. This implies that within the
WKB expansion the centrifugal potential term (\ref{eq:Langerterm2})
should be decomposed as
\begin{equation}
V_{C}(r)= \frac{ L^{2}}{2m r^{2}}+\hbar
\frac{L}{2mr^{2}} \label{eq:Zentrf1}
\end{equation}
where the first term is the classical centrifugal term while the
second term is a quantum correction. Since the WKB expansion proceeds 
in powers of $\hbar$, this latter term has to be treated as a 
perturbation and expanded accordingly. Remarkably, the consequences 
of such a strictly systematic expansion in powers of $\hbar$ seem 
not to have been investigated previously.

We demonstrate that a systematic expansion (SE) about the Kepler problem
yields WKB wave functions that are as accurate as for other potential
problems despite the singularity at $r=0$. Notably, the semiclassical
energy eigenvalues for the hydrogen problem become exact to first
order in $\hbar$ with all higher order corrections vanishing, while
for the problem with LM the exact semiclassical
eigenvalues obtained in lowest order become worse when higher order
corrections are evaluated \cite{Seet}.

We start from the radial Schr\"odinger equation for the hydrogen atom 
\begin{equation}
\left(-\frac{\hbar^{2}}{2m}\frac{d^{2}}{dr^{2}}-
\frac{e^2}{r}+V_{C}(r)\right)\Psi (r)=E \Psi (r)
\label{eq:Schrodrad}
\end{equation}
with $V_{C}(r)$ given by Eq.\ (\ref{eq:Zentrf1}). 
Using the conventional WKB ansatz for the wave function
\begin{equation}
\Psi (r) = 
\exp\left[\frac{i}{\hbar}\sum_{k=0}^{\infty}(-i\hbar)^k S_{k}(r)\right]
\label{eq:Ansatz}
\end{equation}
and expanding in powers of $\hbar$, we obtain for the quantities
\begin{equation}
y_{k}(r,E,L)=\partial S_{k}(r,E,L)/\partial r
\label{eq:defy}
\end{equation}
the recursive set of equations
\begin{eqnarray}
y_{0} & = & p(r,E,L) = \pm
\sqrt{2m(E-V_{\rm eff}(r))}\label{eq:dgl1} \\
y_{1}& = & -\frac{1}{2 y_{0}}\left(y'_{0}
+i\frac{L}{r^2}\right) \label{eq:dgl2}\\
y_{2m} & = &
-\frac{1}{2y_{0}}\left[y_{m}^2+y'_{2m-1}+2\sum_{k=1}^{2m-2}
y_{2m-k}y_{k}\right] \label{eq:dgl3} \\
y_{2m+1} & = &
-\frac{1}{2y_{0}}\left[y'_{2m}+2\sum_{k=1}^{2m-1}y_{2m+1-k}y_{k}\right]
\label{eq:dgl4}
\end{eqnarray}
where
\begin{equation}
V_{\rm eff}(r)= -\frac{e^2}{r}+\frac{L^{2}}{2m r^{2}}.
\label{eq:Zentrf2}
\end{equation}
and where $p(r,E,L)=y_{0}(r,E,L)$ is the classical momentum. Further, the
prime denotes differentiation with respect to $r$.
These equations yield two functions $y^{\left(\pm\right)}(r,E,L)$
depending on the choice of the sign of the momentum $p(r,E,L)$, and
the wave function is a linear combination of the form
\begin{equation}
\Psi\left(r,E,L\right)  =  
 \sum\limits_{\sigma=\pm} c^{(\sigma)}{\rm exp} \left({\frac{i}{\hbar}
\int\limits_{r_{0}}^{r}dr\,
y^{(\sigma)}(r,E,L)}\right) \label{eq:Wfkt1}
\end{equation}
where
\begin{equation}
y(r,E,L)=\sum\limits_{k=0}^{\infty} \left(-i\hbar\right)^k y_{k}(r,E,L).
\end{equation}
The momentum $p(r,E,L)$ has a branch cut which is chosen conveniently
between the classical turning points
\begin{equation}
r_{1,2}=a\left(1\mp\epsilon\right),
\label{eq:umkp}
\end{equation}
where $a$ is the big axis and $\epsilon$ the eccentricity of the
ellipse in the Kepler problem. Dunham \cite{Dunham} has shown that 
by choosing the initial point of
integration $r_{0}$ on the left side of the two classical turning points
and a contour avoiding
the turning points as indicated in Fig.\ 1a, the wave function becomes
\begin{eqnarray}
\Psi\left(r,E,L\right)=\left\{\begin{array}{l@{\ ,\quad}l}
                                c^{(-)}\left(\Psi^{(-)}+\Psi^{(+)}           
                                 \right) &  r_{1}<r<r_{2} \\
                                c^{(-)}\Psi^{(-)} & 
                                                                       
 {\rm elsewhere}\label{eq:Wfkt2}
                                \end{array}\right.
\end{eqnarray}
with
\begin{equation}
\Psi^{(\pm)}(r,E,L)=\exp\left(\frac{i}{\hbar}\int_{r_{0}}^{r}dr\, 
y^{(\pm)}(r,E,L)\right).\label{eq:Wfkt4}
\end{equation}
\vspace{-.5cm}
\begin{figure}
\begin{center}
\leavevmode
\epsfxsize=0.4 \textwidth
\epsfbox{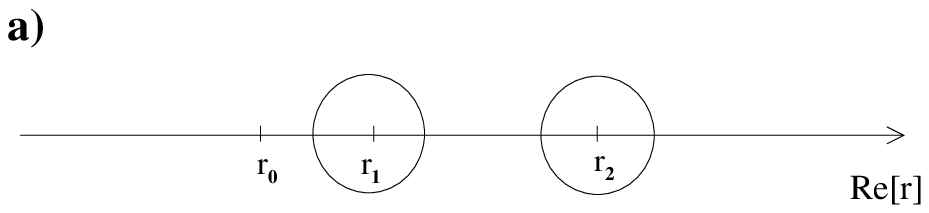}
\end{center}
\end{figure}
\begin{figure}[H]
\vspace{-1cm}
\begin{center}
\leavevmode
\epsfxsize=0.4 \textwidth
\epsfbox{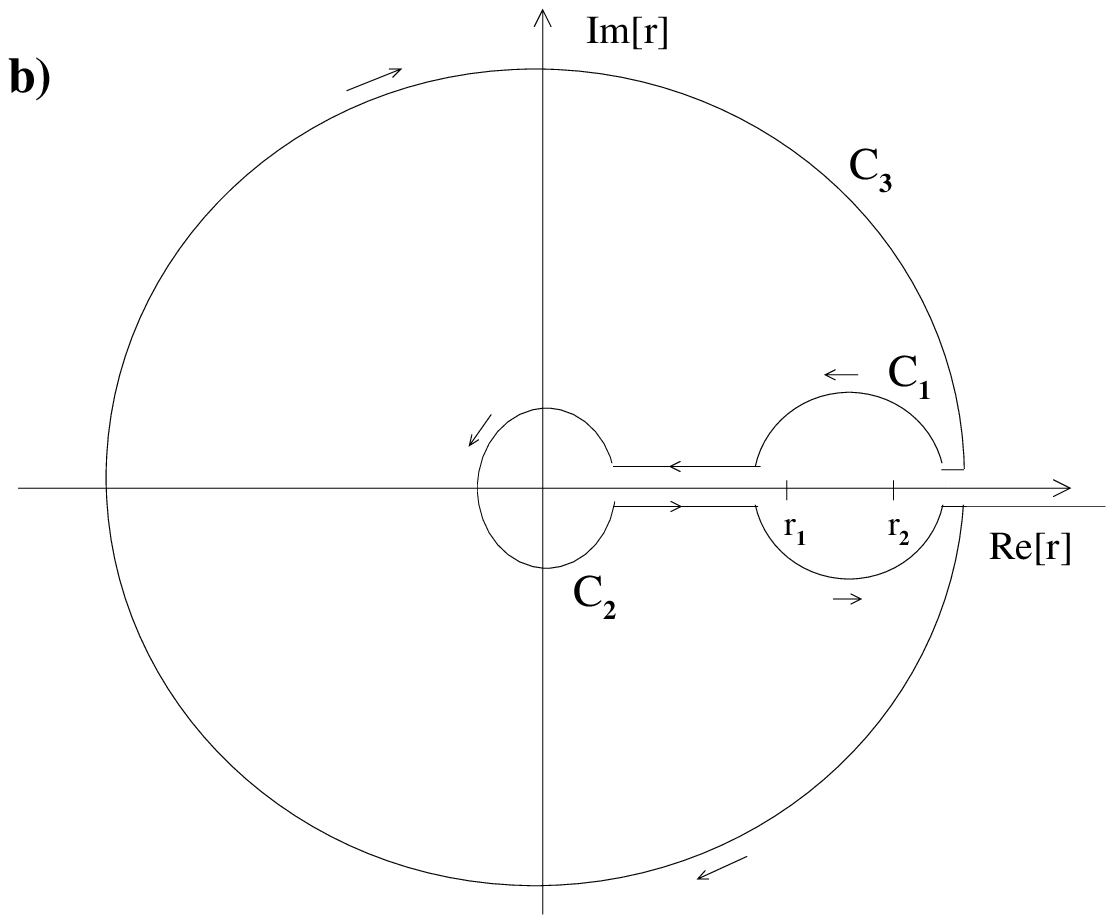}
\end{center}
\caption{ a) The complex $r$ plane with the classical turning points 
$r_{1/2}$. Connecting points
of the classical allowed and forbidden regions one has to avoid the
turning points by integrating along the circles.
 b) Deformation of the integration contour in the complex plane.}
\end{figure}
Since we search for a unique solution, we have to require that the
wave function is independent of whether one integrates above or below
the branch cut. This leads to the condition
\begin{equation}
\frac{i}{\hbar}\oint dr y\left(r,e,L\right)=2\pi i \left(n_{r}+1\right),
\end{equation}
where $n_{r}$ is a positive integer and the integration contour
encircles the branch cut. Using this equation one gets a
quantization of the energy which is related to the 
Bohr-Sommerfeld rule.
To evaluate the contour integrals, we use a technique due to Sommerfeld
which exploits the fact that the $y_{k}(r,E,L)$ have only poles on the
positive real axis.
By this assumption we find
As indicated in Fig.\ 1b one has to
calculate integrals along the contours $C_{2}$ and $C_{3}$
instead of encircling the branch cut. To order $\hbar$ the integrals
are readily evaluated yielding
\begin{displaymath}
\frac{1}{2\pi\hbar}\oint dr\, \left(y_{0}+\frac{\hbar}{i}
y_{1}\right)
=-\frac{L}{\hbar}+\sqrt{-\frac{m
e^{4}}{2 E\hbar^2}}=n_{r}+1,
\end{displaymath}
which gives the exact energy
eigenvalues for the bound states of the hydrogen atom
\begin{equation}
E_{n}=-m e^4/2 \hbar^2 n^2
\label{eq:Rydberg}
\end{equation}
with the principal quantum number $n=n_{r}+l+1$. Corrections of higher
order in $\hbar$ coming from the contour integrals
over the functions $y_{k}$, $k\geq 2$ vanish exactly. To show this we
first investigate the analytical structure of $y_{0}$ and $y_{1}$ at
the origin. We find  
\begin{equation}
y_{0}(r,E,L)=iLr^{-1}+O(r^0)
\end{equation} 
while the power series expansion of $y_{1}$ begins with a linear
term. Consequently the expansion of $y'_{1}$ starts with a constant
term. Now, using
\begin{equation}
y_{2}=-\frac{y_{1}^2+y'_{1}}{2 y_{0}}
\end{equation}
one immediately sees that the expansion of $y_{2}$ begins with a
linear term and therefore the residue of $y_{2}$ at the origin is zero.
Since the recurrence relations (\ref{eq:dgl3}) and (\ref{eq:dgl4})
 contain $y_{0}$ only in the denominator, it is easy to show by
 induction that the Taylor series of all $y_{k}$ with $k \geq 2$ start
 with linear or higher order terms. This implies that the integrals 
along the contour
 $C_{2}$ vanish for all $y_{k}$ with $k\geq 2$. In an analogous way
 one can treat the integrals along the contour $C_{3}$ by replacing
 $r$ by $1/u$ and remembering the additional factor $-1/u^{2}$
 originating from the transformation of the integration measure. One
 finds that the integrals along the contour $C_{3}$ also vanish for
 all $y_{k}$ with $k\geq 2$.
Therefore the semiclassical energy quantization (\ref{eq:Rydberg}) is
exact to all orders in $\hbar$, while in the WKB approximation with Langer
modification higher order terms destroy the exactness of the energy
 eigenvalues.

Next we consider the wave functions. 
Disregarding quadratic and higher powers in $\hbar$ in 
(\ref{eq:Wfkt2}) and (\ref{eq:Wfkt4}),
we arrive at an expression for the  lowest order WKB wave functions
for $r$ on the positive real axis of the form 
\begin{equation}
\Psi(r,E,L)=\frac{1}{2}{\rm
Re}\left[\Psi^{(-)}(r,E,L)+\Psi^{(+)}(r,E,L)\right]
\label{eq:mwwfkt1}
\end{equation} 
with
\begin{equation}
\Psi^{(\pm)}(r,E,L)=\frac{c(E,L)}
{\sqrt{p(r,E,L)}}e^{\pm\left(\frac{i}{\hbar}\int_{r_{1}}^{r}dr\,
p -\frac{i}{2}\varphi-i\frac{\pi}{4}\right)},
\label{eq:mwwfkt2}
\end{equation}
where the additional phase $\varphi(r,E,L)$ arises from
the part of the centrifugal term in (\ref{eq:Zentrf1}) that is linear
in $\hbar$. In fact,
\begin{equation}
\varphi(r,E,L)= -\partial S_{0}(r,E,L)/\partial L
\end{equation}
is just the phase of the classical trajectory in the plane of motion of the
Kepler problem
in terms of which eq.\ (\ref{eq:dgl2}) can be written as
\begin{equation}
y_{1}=-\frac{y'_{0}}{2
y_{0}}-\frac{i}{2}\frac{\partial\varphi}{\partial r}.
\end{equation}
A representation of the WKB wave function as the real part of the
superposition of incoming and outgoing waves as in 
Eq.\ (\ref{eq:mwwfkt1}) was introduced previously
by More and Warren
\cite{M&W} for the standard approach with LM. Since the undesirable
growing part of the wave function has a purely imaginary
coefficient, it is removed when the real part is taken. The normalization
$c(E,L)$ of the wave function is obtained from
\begin{equation}
\frac{1}{2}{\rm Re}\int_{r_{1}}^{r_{2}} dr\,
\Psi^{(+)}\Psi^{(-)}
=\frac{1}{4}\oint dr\, \Psi^{(+)}\Psi^{(-)}=1.\label{eq:norm2}
\end{equation}
which gives
\begin{equation}
c(E)^2 =\frac{2 m}{\pi\hbar^2}\frac{dE}{dn}.
\label{eq:norm3}
\end{equation}
More and Warren refer to the omission of the terms 
$
\Psi^{(+)}\Psi^{(+)}+\Psi^{(-)}\Psi^{(-)}
$
in the normalization integral as
"restricted interference approximation". 
Finally, we get for the WKB wave function of the hydrogen atom in the
classical accessible region between the two turning points
\begin{equation}
\Psi(r,E,L)=\frac{c(E)}{\sqrt{p}}\cos\left(\frac{1}{\hbar}
\int\limits_{r_{1}}^{r}dr\,
p -\frac{\varphi}{2}-\frac{\pi}{4}\right).
\end{equation} 
We now compare the WKB wave functions with the exact ones. A typical 
feature of WKB wave functions is the divergence at the classical turning
points. As can be seen from Fig.\ 2, this behavior is qualitatively 
the same for the Langer modified expansion (LM), our systematic
$\hbar$-expansion WKB(SE), and poor man's WKB(PM) obtained when the
full centrifugal term (\ref{eq:Langerterm2}) is retained in the lowest
order equation. While the WKB(PM) wave function for the ground state
does indeed poorly, the main difference between the WKB(LM) and WKB(SE)
wave functions comes from the fact that the distance between the
turning points of the Langer modified wave functions is smaller. This 
is just a consequence of the shift of the turning points due to the 
LM. Therefore, between the turning points, our wave functions give a
better approximation to the exact ones.
For the s-states, the Langer modified wave functions are constructed 
to vanish at $r=0$ and they have a divergence near the origin since the
left turning point is moved away from $r=0$ by the artificial $1/2$ 
added to the angular momentum number $l$. 
Our wave function doesn't have the right
power law behavior near the origin
but there is only one divergence which is due to the right turning
point. Hence, we see that the
wave functions obtained from a systematic expansion in powers of
$\hbar$ without any {\it ad hoc} manipulation of the hydrogen problem
are at least as accurate as those obtained from the problem with
LM. 
\begin{figure}
\begin{center}
\leavevmode
\epsfxsize=0.4 \textwidth
\epsfbox{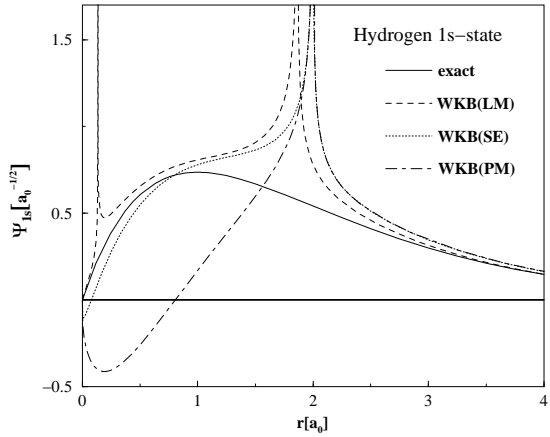}
\end{center}
\caption{WKB wave functions and exact wave functions for the 1s ground
state. The WKB wave functions diverge at the turning points. $r$ is
measured in units of the Bohr radius $a_{0}$.}
\label{fig:fig3}
\end{figure}
\begin{figure}
\begin{center}
\leavevmode
\epsfxsize=0.4 \textwidth
\epsfbox{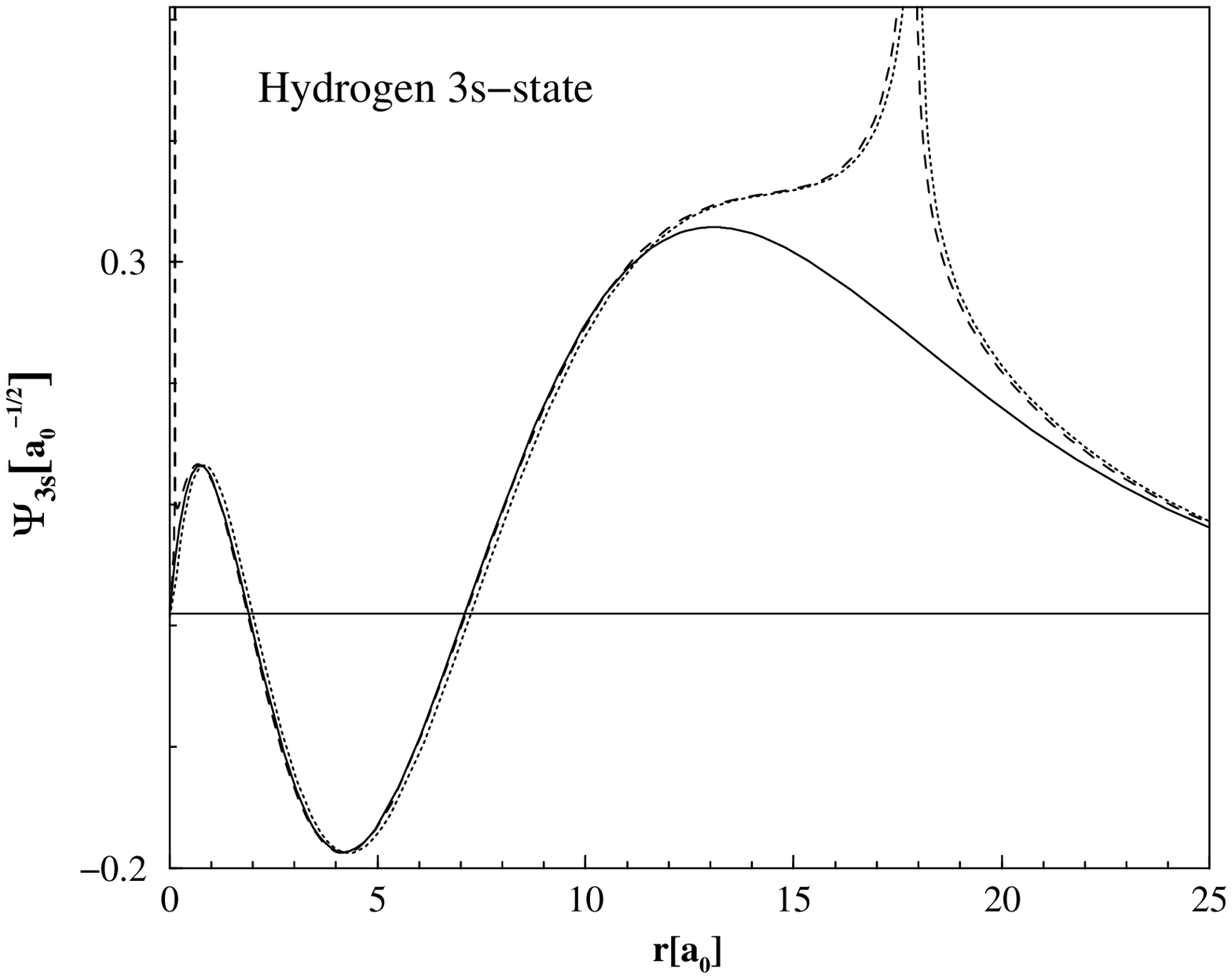}
\end{center}
\vspace{-1cm}
\end{figure} 
 \begin{figure}
\begin{center}
\leavevmode
\epsfxsize=0.4 \textwidth
\epsfbox{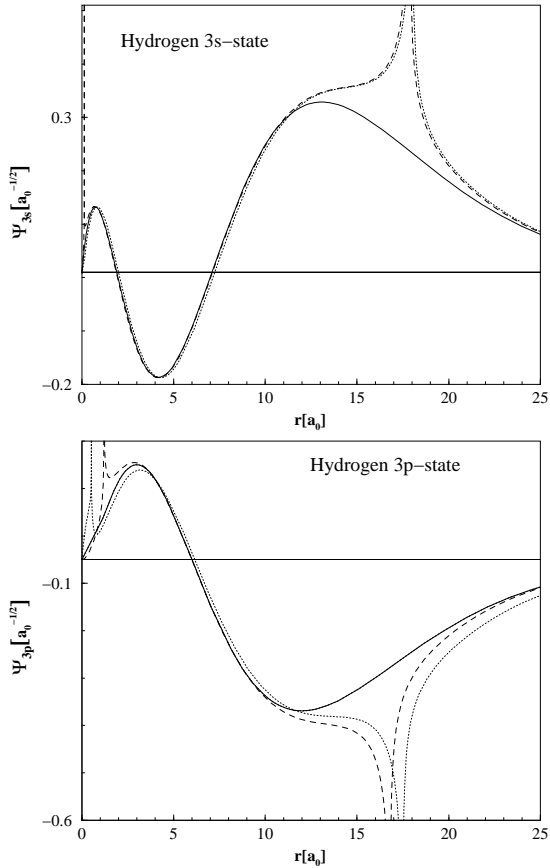}
\end{center}
\caption{WKB wave functions and exact wave functions for the excited 
$3s$ and $3p$ states. Again the full, dashed, and dotted lines denote
the exact, WKB(LM), and the WKB(SE) wave functions, respectively.}
\label{fig:fig4}
\end{figure}

Finally, we calculate radial dipole matrix elements between states
with angular momentum $l$ and $l\pm 1$. Using the restricted
interference approximation, we have
\begin{eqnarray}
&  R^{\pm}_{\Delta n}(E,l) = & \label{eq:matrelm} \\ 
& \frac{1}{4}\oint dr\, \Psi^{(+)}(r,E,L) r
\Psi^{(-)}(r,E+\Delta E,L \pm\hbar). & \nonumber
\end{eqnarray}
Expanding this in powers of $\hbar$, one finds for the leading order term
\begin{eqnarray}
R_{\Delta n}^{\pm\, (0)}(E,l) =
 a_{0}\left(\frac{n^2}{\Delta
n^2}\frac{d}{d\epsilon}J_{\Delta n}(\Delta n\epsilon)\right. \\
 \left. \pm
\frac{n}{\Delta n}\frac{\sqrt{1-\epsilon^2}}{\epsilon}J_{\Delta
n}(\Delta n\epsilon)\right)\nonumber
\end{eqnarray}
where $a_{0}$ is the Bohr radius,
$\epsilon=\left[1-(l/n)^2\right]^{1/2}$ the eccentricity, and
$J_{n}(z)$ a Bessel function.
Naccache \cite{Nac} has obtained this leading order term from the Heisenberg
correspondence principle.
The quantum correction of first order in $\hbar$ is found to read
\begin{eqnarray}
R_{\Delta n}^{\pm\, (1)}(E,l) =  \frac{\Delta
n\omega(E)}{2}\frac{\partial}
{\partial E}R_{\Delta n}^{\pm\, (0)}(E,l)\label{eq:matrelm1Or}\\
 +\frac{1\pm 1}{2}\frac{\partial}{\partial L}
R_{\Delta n}^{\pm \, (0)}(E,l)\nonumber
\end{eqnarray}
with the angular frequency $\omega(E)=\left[-8 E^3/ (m e^6)\right]^{1/2}$ 
of the Kepler problem. In Tab.\ 1 the semiclassical dipole elements
 are compared with the exact ones for some spectral series.
 We note that
for large $n$ and $l$ and small $\Delta n$ the WKB results give rather
 accurate estimates of the exact values. This is expected from a semiclassical
approximation.

In summary, we have shown that a systematic semiclassical expansion of
the hydrogen problem about the Kepler problem yields remarkably
accurate results. In contrast to the common belief no modification of
the WKB expansion is necessary when the centrifugal potential term is
decomposed in the classical centrifugal potential and a quantum
correction. The same method can be employed to other problems with
radial symmetry.

The authors would like to thank Joachim Ankerhold and Phil Pechukas 
for valuable discussions and acknowledge  support by the SFB 276 of
the Deutsche Forschungsgemeinschaft (Bonn). Additional support was 
provided by the Deutscher Akademischer Austausch\-dienst (DAAD).

\end{multicols}
\newpage
\widetext
\begin{table}
\begin{tabular}{|c|c|c|c|c|c|c|c|c|c|}
 n & 2 & 3 & 4 & 5 & 6 & 7 & 8 & 9 & 10 \\ \hline\hline
 1s-np & 1.090(1.290) & 0.512(0.517)  & 0.339(0.306) & 0.257(0.209)
 & 0.208(0.155) & 0.177(0.121) & 0.154(0.098) & 0.137(0.082)
& 0.124(0.069) \\ 
2p-nd & - & 4.542(4.748) & 1.816(1.71) & 1.104(0.975) & 0.802(0.662)
 & 0.641(0.492) & 0.543(0.386) & 0.478(0.314) & 0.432(0.263) \\
4p-ns & - & - & - & 4.673(4.600) & 1.864(1.788) & 1.120(1.044)
 & 0.794(0.718) & 0.614(0.539)
 & 0.501(0.427)   
\end{tabular}
\caption{WKB(SE) dipole matrix elements in units of Bohr's radius
$a_{0}$ and exact quantum mechanical values in parenthesis.}
\label{tab:tab1}
\end{table}
\begin{multicols}{2}

\end{multicols}

\begin{references}
%
\bibitem{Langer}
R.\ Langer, Phys.\ Rev.\ {\bf 51}, 669 (1937).
%
\bibitem{Yi}
H.S.\ Yi, H.R.\ Lee and K.S.\ Sohn, Phys.\ Rev.\ A {\bf 49}, 3277 (1994).
%
\bibitem{Morehead}
J.\ J.\ Morehead, J.\ Math.\ Phys.\  
{\bf 36}, 5431 (1995).
%
\bibitem{Duru}
I.H.\ Duru and H.\ Kleinert, Phys.\ Lett.\ B {\bf 84}, 185 (1979)
%
\bibitem{super}
R.\ Dutt, A.\ Khare, and U.P.\ Sukhatme, Phys.\ Lett.\ B {\bf 181}, 
295 (1986).
%
\bibitem{Friedrich}
H.\ Friedrich and J.\ Trost, Phys.\ Rev.\ A {\bf 54},
1136 (1996), Phys.\ Rev.\ Lett.\ {\bf 76}, 4869 (1996)
%
\bibitem{Seet}
M.\ Seetharaman and S.S.\ Vasan, J.\ Phys.\ 
A {\bf 17} 2485 (1984), have shown how the LM can successively be
adapted to increasing orders of the WKB approximation to preserve the
exact energy eigenvalues.
%
\bibitem{Dunham}
J.L.\ Dunham, Phys.\ Rev.\ {\bf 41}, 713 (1932).
%
\bibitem{M&W}
R.M.\ More, K.H.\ Warren, Ann.\ Phys.\ {\bf 207}, 282 (1991).
%
\bibitem{Nac}
P.F.\ Naccache, J.\ Phys.\ B: Atom.\ Molec.\ Phys.\ {\bf 5}, 1308 (1972).
\end{references}
\end{document}